\documentclass{icrc29}
\usepackage{graphicx,amssymb,amsmath,times}
\begin{document}
%Title of paper
\title[Search for relativistic monopole...]  { Search for relativistic magnetic monopoles with 
the Baikal neutrino telescope NT200 }

\author[Aynutdinov et al ...]  { 
    V. Aynutdinov$^a$, V.Balkanov$^a$, I. Belolaptikov$^g$, N.Budnev$^b$,
    L. Bezrukov$^a$, D. Borschev$^a$,
\newauthor 
    A.Chensky$^b$, I. Danilchenko$^a$, 
    Ya.Davidov$^a$,Zh.-A. Djilkibaev$^a$, G. Domogatsky$^a$, 
\newauthor
A.Dyachok$^b$, 
    S.Fialkovsky$^d$,O.Gaponenko$^a$, O. Gress$^b$, T. Gress$^b$, O.Grishin$^b$, 
    A.Klabukov$^a$, 
\newauthor 
    A.Klimov$^f$, K.Konischev$^g$, A.Koshechkin$^a$, 
    L.Kuzmichev$^c$, V.Kulepov$^d$, 
\newauthor  
    B.Lubsandorzhiev$^a$, S.Mikheyev$^a$, 
    M.Milenin$^d$, R.Mirgazov$^b$, T.Mikolajski$^h$, E.Osipova$^c$,
\newauthor 
    A.Pavlov$^b$,  G.Pan'kov$^b$, L.Pan'kov$^b$, A.Panfilov$^a$,
    \framebox{Yu.Parfenov$^b$}, D.Petuhov$^a$,
\newauthor 
    E.Pliskovsky$^g$, P.Pokhil$^a$, V.Polecshuk$^a$, E.Popova$^c$, V.Prosin$^c$, 
    M.Rozanov$^e$, 
\newauthor 
    V.Rubtzov$^b$, B.Shaibonov$^a$, A.Shirokov$^c$, C.Spiering$^h$, 
    B.Tarashansky$^b$,
 \newauthor 
    R.Vasiliev$^g$, E.Vyatchin$^a$, R.Wischnewski$^h$,
    I.Yashin$^c$, V.Zhukov$^a$ \\ 
 (a) Institute for Nuclear Reseach, Russia\\
 (b) Irkutsk State University,Russia\\
 (c) Skobeltsin Institute of Nuclear Physics, Moscow State University, Russia\\
 (d) Nizni Novgorod State Technical University, Russia\\
 (e) St.Petersburg State Marine Technical University, Russia\\
 (f) Kurchatov Institute, Russia\\
 (g) Joint Institute for Nuclear Research, Dubna, Russia\\
 (h) DESY, Zeuthen, Germany}

\presenter{Presenter: R. Wischnewski (ralf.wischnewski@desy.de), \  
ger-wischnewski-R-abs2-he23-poster}

\maketitle

\begin{abstract}

We present a limit on the flux of relativistic monopoles  obtained
during 994 days of operation of the Baikal neutrino telescope NT200.
The search for relativistic monopoles is based on the enormous amount
of Cherenkov radiation emitted by these particles. 
%This result 
%demonstrates the potential of underwater neutrino telescopes
%with respect to the search for exotic particles.
% ** why potential ? we HAVE already results. potential would be future. **

\end{abstract}

\section{Introduction}

Fast magnetic monopoles with Dirac charge $g=68.5 e$ are attractive 
objects to search for with deep underwater neutrino telescopes.
%Linear   ** What means linear ?? **
The intensity of monopole Cherenkov radiation is $\approx$ 8300
times higher than that of muons. An optical module (OM) of the 
Baikal experiment could detect such an object from a distance up 
%to 60-100 m.
to hundred meters.
Propagating through the Universe, a monopole could be accelerated by magnetic
fields up to energies $10^{21} - 10^{24}$ eV \cite{R.Beck}
,\cite{D.Ryu}. Hence, monopoles with mass below $10^{12}-10^{14}$ GeV
are expected to be relativistic.  The monopole energy loss is about 
10 $\frac{GeV\cdot cm^2}{g}$  for a Lorentz factor $\gamma <10^3$,  
but rises significantly
with energy (by a factor 1000 for $\gamma=10^7$). Monopoles 
with such energy
losses could not cross the Earth. Still, for a wide mass range 
of $10^7 -10^{14}$ GeV  one may search for relativistic 
monopoles from the lower hemisphere, with significant suppression of the
background caused by very energetic atmospheric muons. 

The monopole flux is limited by the Chudakov-Parker bound 
$F_{CP}<$10$^{-15}$ cm$^{-2}$s$^{-1}$sr$^{-1}$ which results from the requirement that 
galactic magnetic fields must be conserved.  The strongest experimental bounds on 
relativistic monopoles ($ \beta > 0.8$) have been obtained by large  
Cherenkov detectors:  Baikal \cite{baikal} and AMANDA \cite{AMANDA}.
   
In this paper we present the result of a search for relativistic  monopoles
with the Baikal 
%deep underwater 
neutrino telescope NT200, based on data taken in the years 1998-2003 (994 days).

\section{Search strategy}

The present stage of the telescope NT200 \cite{baikal_status} takes
data since April, 1998 and consists of 192 optical modules (OMs) at 8
strings. The OMs are grouped in pairs and each pair defines a channel.
A trigger is formed by the requirement of $N_{hit} \ge 4$ fired channels within 500
ns. For such events amplitude and time of all fired channels within time
window 2000 ns are digitized and sent to shore.
The space-time pattern of light recorded from a monopole depends on the
water optical characteristics.
The absorption length of Baikal deep water is $L_{abs}$(480 nm) $ = 20 \div
24 $m and slightly varies during years.  Scattering in Baikal water is
characterized by a strongly anisotropic function $f(\theta)$ with a mean
cosine of the scattering angle $\overline {cos( \theta )}=0.85 \div 0.9$ and
a scattering length $L_s=15 \div 70$ m. 
The OM response to a fast monopole was calculated for a scattering length
30 m.  
OMs facing the monopole path record about one photoelectron at
85 m distance, OM turned away still see one photoelectron at 50 m.
The uncertainty of the scattering length
which varies between 15 m and 30 m  leads to an uncertainty 
of nearly 20$\%$ for 
the effective area for fast monopole registration.     

\begin{figure}[h]
\begin{center}
\includegraphics*[width=0.8 \textwidth,angle=0,clip]{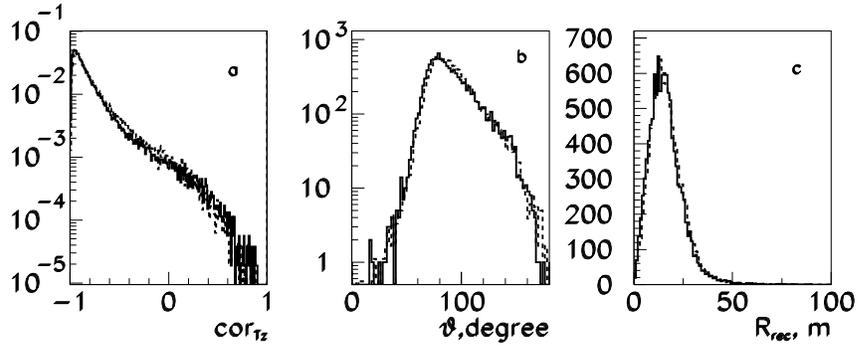}
\caption{\label {mon_3}  (a) $cor_{Tz}$ distribution of experimental
  events with $N_{hit}>30$ (solid) and simulated events from atmospheric 
  muons (dashed);
  (b) Zenith angle distribution  for events selected by cut 0;
  (c) $R_{rec}  $  distribution for events selected by cut 0  }
\end{center}
\end{figure}

The processing chain for fast monopole starts with the selection of
events with a high multiplicity of hit channels.
In order to reduce 
the background  from downward atmospheric muons we
restrict ourself to monopoles coming from the lower hemisphere.    
For an upward going particle the times of hit channels increase with
rising z-coordinates from bottom to top of the detector. To suppress
downward  moving particles, a cut on the value of the z -- time correlation
has been applied:             
 \begin{equation}
cor_{Tz}= \frac{ \sum_{i=1}^{N_{hit}}(t_i- \overline T )(z_i- \overline
  z)} {N_{hit} \sigma_t \sigma_z}>0
\end {equation}
  where $t_i$ and $z_i$ are time and z-coordinate of a fired channel, 
$\overline T$ and $\overline z$ are mean values for times and
  z-coordinates of the event, $ \sigma_t$ and $ \sigma_z$ the root mean
  square errors for time and z-coordinates.
%\begin{center}
%\caption{\label {mon_2}{(a)- $cor_{Tz}$ distribution of experimenal
%    events(solid) with $N_{hit}>30$ and expected distribution from
%    atmospheric muons, (b)- $\chi^2}$  distribution for
%    events sutisfied 0-cut 
%    , (c) Reconstructed zenith angle distribution for the same events
%       (d)- reconstracted $R_{rec}$ distribution    }
%\end{center}
%\end{figure}

The requiments $N_{hit}>30$ and $cor_{Tz}>0$ (cut 0)  
define the first selection stage
and leave 0.015\% of events that satisfy trigger $N_{hit} \ge 6$ on $\ge 3$
strings.
This cut reduces the effective area for monopoles with
$\beta=1$ by almost two times.

The main background for monopole signatures
are muon bundles and single high energy muons. 
The simulation chain of such muons starts with cosmic ray air showers
generated with CORSIKA \cite{corsika}. We use the QGSJET model \cite{qgsjet}
and a primary composition according to \cite{composition}. The MUM
program \cite{mum} is used for muon propagation through water.
In fig.1 we compare 
experimenal data (solid curve) and simulated atmospheric muon events (dashed curve)
with respect to all parameters which have been used
for background rejection. One sees that the simulation describes 
experimental data quite well, even on the level of those rare events.
  
Within 994 days live time, about $ 3 \cdot 10^8$ events with $N_{hit} > 4 $ have been recorded, with
20943 of them satisfying cut 0 ($N_{hit}>30$ and $cor_{Tz}>0$).
All events have been reconstructed according to the standard algorithm \cite{reconstruction}.
For further background supression we use additional
cuts, which essentially reject muon events and at the
same time only slightly reduce the area for registration of relativistic
monopoles.
\begin{enumerate}
\item  $N_{hit}>35$ and $cor_{Tz}>0.4 \div 0.6$
\item  The $\chi ^2$ determined from reconstruction  has to be smaller than 3 ($\chi ^2<3$) 
\item  Reconstructed zenith angle $\theta >100 \deg$
\item  Reconstructed track distance from the center of NT200 $R_{rec} >20 \div 25 $ m.
\end{enumerate}

NT200 took data in various  configurations, due to the different numbers of
temporarily operating channels.  The different cuts on $cor_{Tz}$ and
$R_{rec}$ correspond to different configurations of NT200.
In Table 1 the rejection coefficient, i.e. the ratio  
of events passing cuts 1-4 to those passing cut 0, 
is presented 
(for a subset of configurations representing 50\% of full statistics). 
Also shown is the reduction 
factor for the monopole
effective area $K_A$.    
No events from the experimental sample pass cuts 1-4.

\begin{table}   %[H] add [H] placement to break table across pages
\caption{\label{crflux} Rejection coefficient for experimental
  events and simulated downgoing atmospheric muons, as well as  
  the reduction coefficient ~$K_A$ 
  for the effective area of monopole registration, 
  after application of cuts 1-4}
\begin{center}
\begin{tabular}{||c|c|c|c||} \hline \hline

~Cuts~  & ~EXPERIMENTAL EVENTS~  & ~MC atmospheric muons~  & ~$K_A$
  for monopole~($\beta$=1)      \\

                             \hline \hline

  1        &  0.017 $\underline{+}$1.22 $\cdot$10$^{-3}$& 0.015
                             $\underline{+}$ 8.4 $\cdot$10$^{-3}$ &0.53 \\
                             \hline

  2        & 2.6$\cdot$ 10$^{-3} \underline{+}$
                             4.7$\cdot$10$^{-4}$&3.3$\cdot$10$^{-3} \underline{+}$4.1 $\cdot$10$^{-4}$ & 0.43\\
                             \hline

  3           & 1.1$\cdot$ 10$^{-3} \underline{+}$3.1 $\cdot$ 10$^{-4}$ &
                             1.3$\cdot$ 10$^{-3}\underline{+}$2.5$ \cdot$10$^{-4}$& 0.42\\
                             \hline

  4      & $<$ 2.1 $\cdot$ 10$^{-4}$ 90\% C.L. 
  &2.88 $\cdot$ 10 $^{-4}$$\underline{+}$ 1.2 $\cdot$ 10$^{-4}$&0.38 \\

                             \hline

\end{tabular}
\end{center}
\end{table}

\begin{figure}[h]
\begin{center}
\includegraphics*[width=0.35 \textwidth,angle=0,clip]{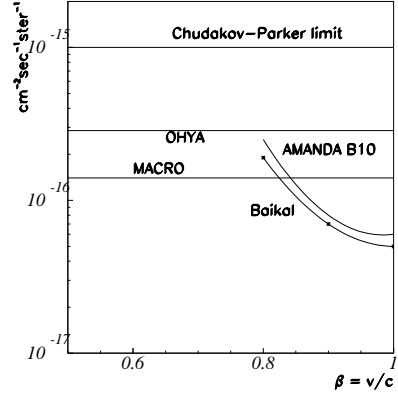}
\caption{\label {mon_02}  $90\%$ C.L. upper limits on the flux of fast monopoles }
\end{center}
\end{figure}

The acceptances $ A_{eff} $ for monopoles with $ \beta=1, 0.9, 0.8 $ have been 
calculated for each configuration of NT200  (17 configurations) 
separately depending on the 
number  operating channels and the concrete values for the rejection cuts.
For the time periods included, it varies between $3 \cdot 10^8$ and $6 \cdot 10^8 $cm$^2$sr 
($ \beta=1$).

From the non-observation of candidate events in NT200 and  the earlier
stages  NT36, NT96 \cite{baikal},  an upper limit on the flux of fast
monopoles on the $90\%$ confidence level is obtained .   The cumulative
acceptances
$A_{eff} \cdot T$ as well as  the $ 90\% $ C.L. upper limits are presented in 
Table 2.

%\begin{figure}[h]
%\begin{center}
%\includegraphics*[width=0.35 \textwidth,angle=0,clip]{mon_02.eps}
%\caption{\label {mon_02}  $90\%$ C.L. upper flux limit     }
%\end{center}
%\end{figure}

 \begin{table}   %[H] add [H] placement to break table across pages
\caption{\label{crflux1} $A_{eff}*T$ and $90\%$ C.L. upper limits on
  the flux of fast monopoles}

\begin{center}
\begin{tabular}{||c|c|c|c||} \hline \hline

        &   $\beta =1$  & $\beta=0.9$  & $\beta=0.8$       \\

                             \hline 

 NT200
  $A_{eff}\cdot T$       &  4.53$ \cdot$ 10$^{16}$  & 3.22$\cdot$ 10$^{16}$ &1.18$\cdot$10$^{16}$ \\
   cm$^2$$\cdot$ sec$\cdot$ sr  &                    &                      &                  \\
                              \hline    
  NT36+NT96 $A_{eff}\cdot T$       & 0.37 $\cdot$ 10$^{16}$ &   0.25
 $ \cdot$  10$^{16}$ & 0.9 $ \cdot$  10$^{15}$\\
    $cm^2\cdot sec\cdot sr$        &                        &            &         \\
                         \hline
  90 \% C.L. upper flux limit     & 0.5 $\cdot$ 10$^{-16} $ & 0.7
 $ \cdot$ 10$^{-16}$&1.92 $\cdot$ 10$^{-16}$\\ 
 cm$^{-2}$$\cdot$ sec$^{-1}$$ \cdot$ sr$^{-1}$      &    &     &           \\
                                   \hline

\end{tabular}
\end{center}
\end{table}

     In fig.2, the 90\% C.L. upper limit  obtained with the Baikal neutrino
     telescope  for an isotropic flux of fast monopoles is compared
     to the final limits from the undergrund experiments Ohya \cite{ohya}
     and MACRO\cite{macro} and to the published limit of the underice
     detector AMANDA\cite {AMANDA}.

   \section{Acknowledgements}

    This work was supported by the Russian Ministry of Education and
    Science, the German Ministry of Education and Research and the
    Russian Fund of Basic Research (grant 05-02-17476, 04-02-17289 and
    02-07-90293) and by the the Grant of the President of Russia NSh-1828.2003.2

\end{document}